# Status Quo, Critical Reflection, and the Road Ahead of Digital Nudging in Information Systems Research: A Discussion with Markus Weinmann and Alexey Voinov

Christian Meske
*Freie Universität Berlin*, christian.meske@fu-berlin.de

Ireti Amojo
*Freie Universität Berlin*

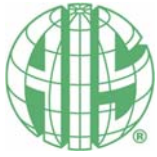

# Communications of the Association for Information Systems

# Status Quo, Critical Reflection, and the Road Ahead of Digital Nudging in Information Systems Research: A Discussion with Markus Weinmann and Alexey Voinov

**Christian Meske**
Department of Information Systems
Freie Universität Berlin and
Einstein Center Digital Future
christian.meske@fu-berlin.de

**Ireti Amojo**
Department of Information Systems
Freie Universität Berlin and
Einstein Center Digital Future

**Abstract:**

Research on digital nudging has become increasingly popular in the information systems (IS) community. In this paper, we overview the current progress of, critically reflect on, and discuss further research on digital nudging in IS. To do so, we reviewed the literature and interviewed Markus Weinmann from Rotterdam School of Management at Erasmus University, one of the first scholars who introduced digital nudging to the IS community, and Alexey Voinov, Director of the Centre on Persuasive Systems for Wise Adaptive Living at University of Technology Sydney. We uncovered a gap between what we know about what constitutes digital nudging and how we can actually put consequent requirements into practice. In this context, the original nudging concept bears inherent challenges about, for example, the focus on individuals' welfare, which, thus, also apply to digital nudging. Moreover, we need to better understand how nudging in digital choice environments differs from that in the offline world. To further distinguish itself from other fields that already tested various nudges in many different domains, digital nudging research in IS may benefit from a design science perspective in order to go beyond testing effectiveness and provide specific design principles for the different types of digital nudges.

# 1 Introduction

According to traditional theories of economics and management science and as the term "homo economicus" reflects, humans are rational beings (Brzezicka & Wisniewski, 2014). However, researchers have discussed new theories and perspectives to uncover the real "conditio humana" (Dierksmeier, 2015) and, thus, to better understand human behavior in all its facets. In this context, Richard Thaler, Nobel Prize winner and one of the most prominent scientists of behavioral economics, combined psychological decision-making theories at the micro level with collective behavior at the macro level.

His work builds on three essential cornerstones that explain irrationalities in human behavior. First, individuals may have varying levels of self-control. This aspect refers to situations in which agents cannot carry out their optimal plans even if they can compute them (e.g., Thaler & Shefrin, 1981). Second, individuals may have "social preferences" and, hence, a strong tendency to pay attention to others' gains or welfare rather than pure selfishness as the homo economicus would suggest. For instance, Thaler showed how the perceived fairness affects individuals' behavior in consumer and labor markets, which has significant implications for optimal firm behavior (e.g., Kahneman, Knetsch, & Thaler, 1986). Third, individuals may have a "bounded rationality". Bounded rationality refers to the assumption that humans have two interconnected cognitive systems: unreflective and automatic (System 1) and reflective and slow (System 2) (Kahneman, 2003; Thaler & Sunstein, 2008). Researchers believe System 1 thinking to be the most regularly cognitive process in everyday life based on cognitive heuristics and biases (Stanovich & West, 2000; Kahneman, 2003). The brain handles these systems separately, which means that System 1 provides premises and contextual information (e.g., hold breath under water), and enables System 2 to process information (control body activity) (Hansen & Jespersen 2013). The psychologist Fogg (2003) has discussed the concept of bounded rationality and its consequences on human decision making. He defined persuasive mechanisms in virtual environments (persuasive technologies) as "any interactive computer system designed to change people's attitudes or behaviors" (p. 1). Given the broad perspective on persuasive technologies that this definition provides, there has been an ongoing debate on the demarcation aspects of Digital Nudging in contrast to persuasion.

In 2008, based on their developed understanding of human decision making, Thaler and Sunstein published the concept of "nudging", which has ever since earned a lot of attention in politics, law, social science, and various other fields. According to this concept, a certain choice architecture surrounds every person. Each person willfully creates this architecture or it emerges at random, and it represents an individual's immediate environment that directly influences their decision making. In consequence, the authors assume that altering the choice architecture will consequently lead to a change in decision making. Thus, in referring to "nudging", Thaler and Sunstein refer to influencing humans through altering their immediate environments to increase their long-run welfare as they themselves judge it. Thus, nudging represents a form of libertarian paternalism and includes the requirement that a nudge does not restrict the choices available to an individual (Cohen, 2013).

In recent years, the nudging concept has also become an important research focus in the information systems (IS) community, though IS researchers have mainly termed it "digital nudging" (e.g., Weinmann, Schneider, & vom Brocke, 2016; Hummel, Schacht, & Maedche, 2017; Meske & Potthoff, 2017; Schneider, Weinmann, & vom Brocke, 2018; Kretzer & Maedche, 2018; Hummel & Maedche, 2019). In this paper, we summarize the progress in this young research area and overview existing groundwork. In particular, we focus on discussing varying definitions, design models, and first empirical findings. Furthermore, we discuss and critically reflect on these first steps that digital nudging research has taken. Finally, we conclude the paper by discussing potential future research.

We overview the existing literature on digital nudging to serve as the basis for our discussion with Markus Weinmann from Rotterdam School of Management at Erasmus University and Alexey Voinov from the University of Technology Sydney (UTS). Markus was one of the first researchers to introduce digital nudging to the IS community in 2016. Alexey is well known for his work in participatory modeling and is director of the new Centre on Persuasive Systems for Wise Adaptive Living (PERSWADE) at UTS.

The paper proceeds as follows: in Section 2, we present the current progress on digital nudging in IS scholarship. We also introduce important definitions, present currently discussed design approaches, summarize empirical results on digital nudging, and, hence, discuss the effectiveness of digital nudging in different domains. In Section 3, we present the transcript for our interview with Markus Weinmann and Alexey Voinov. In Section 4, we discuss the findings and interviews. Finally, in Section 5, we conclude the paper.

## 2 Status Quo of Digital Nudging in Information Systems Research

In this section, we present digital nudging research's current progress based on analyzing the literature. To do so, we follow Paré, Trudel, Jaan, and Kitsiou's (2015) typology for literature reviews that researchers have commonly applied in IS research. In reviewing the literature, we focus on 1) summarizing prior knowledge by applying a comprehensive search strategy and 2) using both conceptual and empirical primary sources. Therefore, one can characterize our review as a scoping review (Paré et al., 2015) in which we map digital nudging research in IS around three themes: definitions, design models, and effectiveness. However, we use the results as the basis for interview with Markus Weinmann and Alexey Voinov. Following the review and interview, we critically reflect on our findings. Therefore, the literature analysis eventually becomes attributes of a critical review (Paré et al., 2015).

Many publications outside the IS field have investigated the validity of human-behavior interventions in different offline and online settings and domains (see, e.g., Fogg, 2003; Choe, Jung, Lee, & Fisher, 2013; Almuhimedi et al., 2015; Acquisti et al., 2017). As such, research has quite intensively analyzed the effect that potential nudges (or interventions) have on human behavior. Such effort raises questions about how IS research can originally contribute to knowledge in this area. However, to reflect on the advancements that research on digital nudging in IS has made, we first need to establish a status quo of corresponding research efforts in the community. Accordingly, in our literature review, we mainly focus on publications from IS conferences and journals using AIS electronic Library (AISeL) and Web of Science to provide and discuss the status quo. To select relevant work, we searched for publications that explicitly focused on nudging and referred to that term.

### 2.1 Definitions, Basic Principles, and Ethics

Before we dive into the IS literature, we briefly discuss the basic definition for nudging and how IS research has understood it. Thaler and Sunstein (2008), the scholars who first proposed the nudging concept, defined a nudge as "any aspect of the choice architecture that alters people's behavior in a predictable way without forbidding any options or significantly changing their economic incentives" (p. 6). Further, they postulated that "to count as a mere nudge, the intervention must be easy and cheap to avoid (as) nudges are not mandates" (p. 6) and that nudgers must try "to influence choices in a way that will make choosers better off, as judged by themselves" (p. 5). Most subsequent definitions for nudging have adopted Thaler and Sunstein's (2008) definition or only slightly changed the terminology. Arguing from a more specific and less generalist perspective, Hansen (2016) defined a nudge as:

> *any attempt at influencing people's judgment, choice or behavior in a predictable way, that is (1) made possible because of cognitive boundaries, biases, routines, and habits in individual and social decision-making posing barriers for people to perform rationally in their own self-declared interests, and which (2) works by making use of those boundaries, biases, routines, and habits as integral parts of such attempts.* (p. 4)

In regards to nudges' libertarian paternalist characteristic, Hansen (2016) further stated that behavioral influence should avoid "1) forbidding or adding any rationally relevant choice options, 2) changing incentives, whether regarded in terms of time, trouble, social sanctions, economic and so forth, or 3) the provision of factual information and rational argumentation" (p. 4).

Recently, IS research has also picked up the nudging concept and modified its definition to fit new contexts of digital environments. In particular, Weinmann et al. (2016) introduced the term "digital nudging" to a broader IS audience and referred to it as using "user-interface design elements to guide people's behavior in digital choice environments" (p. 1). As designers or choice architects (here: nudgers) always construct virtual environments, user interfaces represent the virtual contact point through which they influence users' decisions (Weinmann et al., 2016). Digital choice environments include websites, mobile apps, and enterprise resource planning (ERP) or customer relationship management (CRM) systems in various domains, from e-government to e-commerce, in our private and professional lives. Exemplary nudges include incentives, defaults, feedback, or structured complex choices (Weinmann et al., 2016). Moving to a more specific definition, Meske and Potthoff (2017) defined digital nudging as "a subtle form of using design, information, and interaction elements to guide user behavior in digital environments, without restricting the individual's freedom of choice" (p. 2589). Mirsch, Jung, and Lehrer (2018) also defined digital nudging in a way that links closely to previously proposed definitions for nudging in defining it as the "attempt to influence decision-making, judgment, or behavior in a predictable way by counteracting the cognitive boundaries, biases, routines, and habits that hinder individuals from acting to their own benefit in the digital sphere" (p. 3). Recently, Lembcke, Engelbrecht, Brendel, Herrenkind, and Kolbe (2019b) have added more specificity to digital nudges compared to initial definitions in defining them as a "any intended and goal-oriented intervention element (e.g., design, information or interaction elements) in digital or blended environments attempting to influence people's judgment, choice, or behavior in a predictable way" (p. 10).

As the definitions show, digital nudging at its core always concerns influencing (or manipulating) human behavior, which raises some ethical questions. In this context, van den Hoven stated in an interview:

> *Via their designs for systems and artifacts they come to have an incredible impact on the lives of others: cables, code, search and reach algorithms, standards, ontologies, authorization matrices, menus, voting procedures, aggregation mechanisms, recommender systems, reputation systems. …It will become more and more important in the future to be able to design systematically for moral, legal and social requirements.* (Maedche, 2017)

However, while researchers have provided many definitions to transfer the nudging approach to the digital (choice) environment, they have seldom discussed basic ethical or moral considerations in the underlying concept. The topic has only recently begun to receive more attention. For instance, based on existing literature, Lembcke et al. (2019b) discuss how much effort individuals should justifiably have to make to preserve their freedom of choice, how concealed a nudge may become such that one could still consider it transparent, or how aligned choice architects' goals need to be with individuals' goals in order to render a nudge as justifiable. Yet, while these considerations are important, we still lack concrete ethical guidelines for researchers and practitioners.

## 2.2 Design Models

The following research on digital nudging in IS research has provided guidelines or strategies for designing and implementing nudges in virtual environments. Meske and Potthoff (2017) formulated the following basic stages that nudge designers should consider: 1) analyze the target audience, reasons for nudging, and goals; 2) design adequate nudging elements (i.e., default-setting, priming, or reminders) and consider relevant factors (i.e., nudgees' ability, motivation, or context); and 3) evaluate the nudge's design while considering the degree to which it adheres to libertarian paternalist principles (freedom of choice, respecting of user preferences, unchanged incentives) (Meske & Potthoff, 2017). In addition, the authors list and describe potential nudges, such as anchoring, customized information (tailoring), decision staging (tunneling), default setting, framing, informing, limited time window, praise, precommitment, priming, reminders, simplification and social influence. Similar, Mirsch, Lehrer, and Jung (2017) systematically reviewed the literature to identify potential nudge mechanisms. Mirsch et al. (2017) also provided a process model based on Weinmann et al.'s (2016) work with the following steps: define (context and goals), diagnose (decision process to determine relevant psychological effects), select (appropriate nudges to alter behavior), implement (design of nudges and choice architecture) and measure (evaluation of nudges).

Similar to Meske and Potthoff (2017) and Mirsch et al. (2017), Schneider et al. (2018) highlight the importance for designers to understand and adequately address underlying cognitive heuristics and biases when developing nudging strategies. In addition to providing an experimental setting in which they illustrate said heuristics and biases during online decision making, they also provide a stepwise approach (design cycle) for a nudge design strategy (Schneider et al., 2018): 1) define a goal for nudging, 2) understand the nudgees (users), 3) actually design the nudge with respect to heuristics and biases, and 4) iteratively testing the nudge to assess its effectiveness (Schneider et al., 2018).

Building on previous nudge design models and knowledge about their underlying psychological effects, Mirsch et al. (2018) provide an evaluated and to date the most detailed approach for designing digital nudges. They used a design science research approach to derive what they call a "digital nudge design method" that comprises: 1) the digital nudge context, 2) the digital nudge ideation and design, () the digital nudge implementation, and 4) the digital nudge evaluation (Mirsch et al., 2018). The originality of their contribution lies in their describing tools and techniques such as considering qualitative (e.g., interviews) and quantitative instruments (e.g., user surveys) to complement and improve how one designs a nudge during the respective phases.

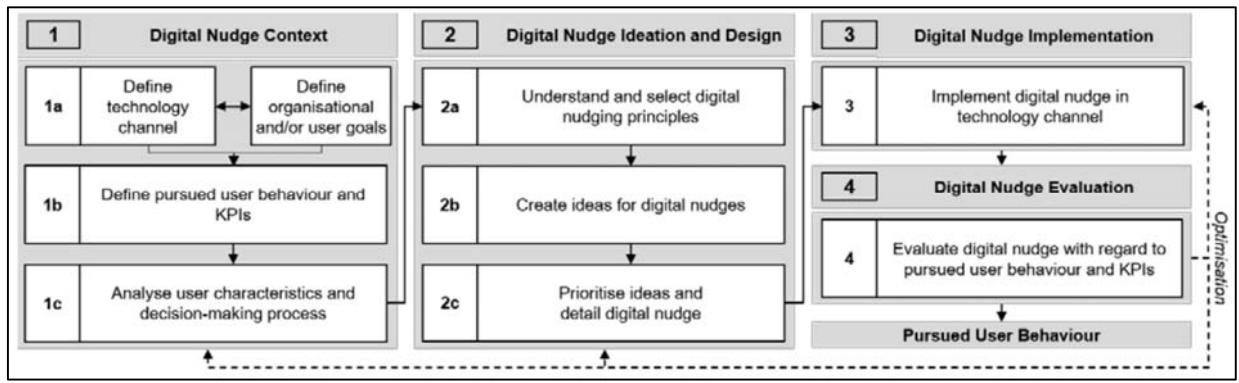

**Figure 1. Digital Nudge Design Method (Mirsch et al., 2018, p. 10)**

Other research in this regard has set on developing guidelines for future design science research approaches to test existing design methods and specifically evaluate the effects of nudging mechanisms (e.g., Eigenbrod & Janson, 2018), such as their effect in comparison to other effects that influence choice in online environments (e.g., Djurica & Figl, 2017).

## 2.3 Effectiveness

Empirical research on digital nudging in IS research revolved around the hidden potentials and general opportunities for nudging in the field's various research subfields, such as data security, Internet privacy, e-commerce decisions, and social media. For instance, Huang, Chen, Hong, and Wu (2018) the influence of social sharing nudges on an online education and career information platform in China. They assessed that social sharing through using relational (or social) capital messages (i.e., "Share this webpage with your friends! They may find the information helpful!") more significantly affected social sharing than nudges that, for instance, plainly requested users to share information (i.e. "Share this webpage with your friends!"). Schneider, Lins, Grupp, Benlian, and Sunyaev (2017) conducted further research on the use and influence of communication arguments as a form of nudging. They argued that the combination of claims and supporting arguments that users provide on online platforms may address, for example, privacy concerns and, consequently, guide user behavior in digital choice environments. Accordingly, they assessed arguments framed as either promoting online verification or preventing privacy concerns. They found, that even though the prevention focus had stronger effects, both arguments significantly increased the conversion rates of online verifications only if they also included supporting data (here, assurance seals) (Schneider et al., 2017). In their study, Wang, Zhang, and Hann (2014) found that reminding users about their online audience when disclosing information most effectively nudged users to reduce their online disclosure. Likewise, nudges in the form of alerts that sensitize users about the collected data during their app use similarly increase security and privacy decisions (Almuhimedi et al., 2014). Cao, Hui and Hong (2018) considered nudging in terms of information that peers disclose. They showed that nudging users to reflect on the privacy consequences of the disclosed information online increases privacy in online communities (among other things). In public social media, Kroll and Stieglitz (2019) found that privacy-related digital nudges show effects of (perceived) control, (perceived privacy) risk, and trust (in the provider) on self-disclosure in social networks (here, Facebook). Meanwhile, Choe et al. (2013) assessed framing nudges in their work and found that implementing nudges to influence users' perceived trust of mobile apps had little effects.

Wang et al. (2018) reaffirmed the positive social correlation between online word-of-mouth nudges and online product ratings. By using a design science research approach, Rodriguez, Piccoli, and Bartosiak (2019) considered digital nudging mechanisms (e.g., default trigger) to design an artefact to reduce how much students procrastinated. Other research on the influence of default-setting nudges has also showed that default-settings, such as revealing information about payment options and their effects on the environment, have a positive influence on nudging individuals to make more environmentally sound options during online flight booking processes (Székely, Weinmann, & vom Brocke, 2016). Also, information nudges about eligibility and access to fee waivers showed a significant increase in applications in low income immigrants (Hotard, Lawrence, Laitin, & Hainmueller, 2019). Moreover, Schneider et al. (2018) focused on online user engagement in the context of a crowdfunding platform based on modifying the content or choice architecture pertaining to the following three cognitive heuristics and biases: decoy effects, scarcity effects, and middle-option biases. They found that, even though all heuristics had an effect on user engagement, the importance of finding adequate nudges to eventually affect targeted heuristics constitutes an integral part of applied design strategies (Schneider et al., 2018).

In a descriptive study, Özdemir (2019) holistically overviewed how researchers have applied nudging in a variety of digital interfaces (e.g. mobile, social media, or enterprise systems) to lay out concepts, patterns, and the usage of nudges in virtual environments. Hummel, Toreini, and Maedche (2018) based their research on the assumption that nudging can sometimes be less effective than expected. Therefore, by using eye-tracking technology in their experimental study, they assessed whether interactive real-time feedback can help individuals recognize digital nudges and, hence, the latter's effectiveness (Hummel et al., 2018). Finally, in a systematic literature review, Hummel and Maedche (2019) built on the idea of assessing the (in)effectiveness of nudges more closely. Despite showing that digital nudging can be less effective than assumed, they identified possibly reasons for nudges' ineffectiveness and provided a morphological box to help both scholars and practitioners holistically understand digital nudging (Hummel & Maedche, 2019).

In enterprises, Klesel, Jahn, Müll, and Niehaves (2016) conducted the first research on possible nudging tools' design principles. They used nudging theory to design and introduce mechanisms, which, contrary to failed enforcement strategies, could support employees' work discontinuance after regular working hours and, thus, improve their work-life balance. Accordingly, the researchers identified default, social, or reminder nudges as among 14 identified ways to operationalize nudges in enterprise settings (Klessel et al., 2016). Moreover, Stieglitz, Potthoff, and Kißmer (2017) show that nudging can have positive influences during enterprise technology-adoption processes. Also, in enterprise research, Kretzer and Maedche (2018) found that their pre-defined social nudges (cohesion, business function, geographical distance, and hierarchy) had a positive influence on users' decision-making (here: choosing report recommendations).

Thus far, digital nudging research has not yet discussed digital nudging's cultural aspects. While some research acknowledges that nudging has different effects for people from different socio-cultural groups (Morgan, Deedat, & Kenten, 2015), researchers have yet to thoroughly scientifically debate the role that culture plays and the possible ineffectiveness of certain digital nudges in countries with more non-libertarian paternalist policies.

## 3 Interview

In this section, we discuss digital nudging research's progress with Markus Weinmann and Alexey Voinov, reflect on its advancements and the concept itself, and discuss further research.

Markus Weinmann is Associate Professor at the University of Rotterdam. He holds a doctorate from the Technical University of Braunschweig and has been a visiting researcher at several universities, including City University of Hong Kong and Queensland University of Technology. His research concerns "IT and Behavioral Economics" and, in particular, how people judge, decide, and behave in online environments. Areas of application include crowdfunding (e.g., Kickstarter), crowdsourcing (e.g., Transfermarkt), and online ratings (e.g., TripAdvisor). His research has appeared in leading journals (e.g., *MIS Quarterly*), won prestigious awards (e.g., European Research Paper of the Year), and been featured in premier media outlets (e.g., *The Wall Street Journal* and *Der Spiegel*).

Alexey Voinov is a Distinguished Professor at the School of Information, Systems and Modeling at Faculty of Engineering and IT of University of Technology Sydney. He is leading the faculty Centre on Persuasive Systems for Wise Adaptive Living (PERSWADE). Before joining UTS, Alexey was professor of Spatio-Temporal Systems Modeling for Sustainability Science at ITC, University of Twente, and coordinated the Chesapeake Research Consortium Community Modeling Program. For over ten years, he worked with the Institute for Ecological Economics, the University of Marylan, and University of Vermont. His main interests fall in integrated, participatory modeling, which includes merging qualitative and quantitative methods with behavioral science to find better ways to communicate scientific findings and turn them into actions.

We interviewed these two scholars via Skype (video) on 26 April, 2019. We taped and transcribed the spoken conversation for this paper. The transcription contains all questions and answers. While we created an initial set of questions beforehand, other questions emerged during the course of the interview. We exclude non-lexical vocalizations such as "um" and pauses from the transcript. We have also modified literature that the interviewees mentioned to adhere to the journal's guidelines and improve readability. Both authors checked and edited the text. Eventually, we asked Markus Weinmann and Alexey Voinov if the final transcript still reflected their original statements, and both fully confirmed that it did.

### 3.1 Introduction

**CM:** **Markus, what gave you and your colleagues the idea to write the catchword on digital nudging in 2016 that was one of the first IS publications in that context?**

**MW:** I have always been interested in the field of judgment and decision making. The ideas of Richard Thaler and Cass Sunstein, described in their bestseller "Nudge: Improving Decisions about Health, Wealth, and Happiness", caught my interest particularly; they suggest that the design of our decision-making environment has an enormous influence on our decision-making behavior. Information Systems deals with the design of systems and decisions, so we, that is Christoph Schneider from the City University of Hong Kong and Jan vom Brocke from the University of Liechtenstein, found using "nudging" in the digital context an obvious extension.

**CM:** **Alexey, you are director of PERSWADE. Why is digital nudging of your interest?**

**AV:** PERSWADE Center has been largely created out of frustration. In so many ways, science today knows so much more than it is able to communicate and implement in action. The climate change saga is one such example. We have reached the point when school kids seem to be more understanding than politicians. There are many more examples. Biodiversity, resource depletion, pollution, economic growth—we can go on and on with examples where science has been quite certain about what our actions should be, yet we are failing miserably to take those actions.

So, we have formulated the goal of the center as "developing and applying persuasive technologies and system science for social innovation that can help humanity to move toward sustainable, wise, adaptive living, realizing that our well-being and prosperity are very much dependent upon the Earth life-support system".

People can be motivated to do things. From history and from everyday practice, we see how people are persuaded to change their behavior in ways that suit certain interests, groups or parties. Rarely, we see these persuasion techniques applied for the benefit of the society as a whole, for sustaining life on this planet. Consumerism is an example of successful social engineering that was essential for the success of capitalism.

We would like to see how similar or other methods as digital nudging can be used to turn the course, to steer away from the imminent disaster, or, at least to prepare, to adapt in ways that would be beneficial for the majority of the population, not just for the selected few.

### 3.2 Current Definitions and Design Models for Digital Nudging

**CM:** **While I understand the motivation, the goal to elicit a certain user behavior does yet not seem to be something new and could be understood as an integral part of every system design approach. What kind of originality do you think does nudging or digital nudging bring into the information systems community? What has been missing?**

**MW:** Certainly, one goal in information systems is to induce certain user behaviors, I agree. But the question is how to achieve this goal. One method is digital nudging, which relies on designing the decision environment like apps or websites in such a way that users change their behavior relatively automatically and unconsciously. For this purpose, IS scholars transfer theories and findings that have long been known in psychology and behavioral economics to applications in information systems. Thus, nudging helps to improve the design of information systems, backed by evidence-based research.

**CM:** **This may be true, but in that case, how does nudging differ from persuasion? Compared to persuasion and persuasive technology, is digital nudging something new or just "old wine in new bottles"?**

**AV:** Digital nudging appears to me as a type, a subset of persuasion. Persuasion can be aggressive, persistent, intensive. Nudging seems to be a softer type of persuasion, which just chases you in various circumstances, but does not really demand anything from you. It is softer, it is more on the awareness building and informing side, rather than direct luring, enticing, requiring or threatening.

**MW**: The two concepts have in common that they seek behavior change, but how they do so differs. Whereas persuasion seeks to change attitudes (often by changing minds) and thereby behavior, nudging seeks to change the decision environment (by changing the contexts). When nudged, people change their behavior often relatively unconsciously and automatically. I think, the two concepts complement each other well, and their effectiveness certainly depends on the decision situation (for example, its complexity) and the decision maker (for example, the willingness to invest cognitive effort). However, in many situations, it may be easier to change a situation's context through nudging than to change people's minds through persuasion.

**AV**: Persuasion does not really care how behavior will be changed—directly or indirectly. As long as the change is attained, the persuasion was successful. That is why persuasion would be happy to include nudging as a tool. Moreover, it is sometimes pretty hard to distinguish between changing contexts or minds. For example, what does conventional advertising do? Does it change the mind or the context? I would say—both. If you go along the street that is all decorated by images of Coca-Cola bottles—that seems to be pretty contextual, despite the fact that this can be quite aggressive and annoying, yet effective.

**MW**: I only partially agree. Persuasion is defined as "the act of causing people to do or believe something" (Merriam-Webster). A synonym is convincing; in other words, to change someone's attitude or mind. In contrast, nudging is defined as "any aspect of the choice architecture that alters people's behavior in a predictable way without forbidding any options", so nudging is not about convincing but a clever design of the decision environment that makes people choose subconsciously and automatically. Therefore, I think there are differences between the two concepts, and I see nudging not as part of persuasion but as complementary.

**AV**: I agree that they are not exactly the same. I just suggest that nudging is a subset of persuasion. Just like a cucumber is a vegetable, it does not mean that all vegetables are cucumbers. It's also in the definition of persuasion, where you have an "or": do or believe. So persuasion covers doing without believing, which then includes the definition of nudging. What bothers me is that if we are expecting somebody to do something without even been convinced, without realizing that they are doing it, isn't that akin to manipulation?

**CM: I will come back to the aspect of manipulation later as I would like to stick with the basic understandings on digital nudging first. In this regard, how do you two evaluate the current definitions of digital nudging? Do you agree on the existing propositions or would you change, or add something based on the current advancements in the field?**

**MW:** Definitions are often subject to an accuracy-simplicity tradeoff: short definitions are simple but tend to lack precision, and long definitions tend to be precise but complex. When we first defined "digital nudging" in 2016, we wanted to keep it short, proposing that "digital nudging is the use of user-interface design elements to guide people's behavior in digital choice environments". Only later in the same article did we also discuss ethical components like freedom of choice.

Today, I would add to the definition that ethical questions must be taken into account: for example, "and should be used to help people make better choices while maintaining their freedom of choice". Strictly speaking, this addition does not refer to the digital aspect but to nudging in general. Some scholars have provided even more comprehensive definitions, most of which I agree with, such as Meske and Potthoff (2017) and Mirsch et al. (2017).

**AV:** I am not sure to what extent my definition above is mainstream for nudging in general and digital, in particular. Probably not quite. So, I guess I disagree and think that it can be changed. I am not especially comfortable with all the rhetoric about how nudging is 'soft', non-intrusive, with no obstruction of free choice, etc. What is the goal of Digital Nudging after all? We do it for a purpose, right? And the purpose is to change something in human behavior, in their free choice. So, either we change it or not. What is the purpose of pretending that nudging is not restricting freedom of choice, yet it is changing it? You cannot be a little pregnant.

**MW:** Interesting point. Indeed, the goal of nudging is to change behavior (very similar to persuasion techniques), but not by prohibiting people and restricting their freedom of choice, but by adapting the presentation of choices—the choice architecture—and preserving their freedom of choice. An example: to prevent people from choosing a particular option, I could either not present that option at all (restricting the freedom of choice), or leave the choice set open, but set a default (preserving the freedom of choice).

**AV:** But that is where we already embark on the delicate track of extracting the desired result from people knowingly orchestrating the choice process based on what we know about human psychology and pretending that they are free to make the choice, while we know that this freedom is actually restricted.

**CM:** **Your different perspectives on the basic conception of digital nudging may also have an influence on the construction of such nudges. Do you think we have a sufficient understanding on how to design such artefacts? For instance, how do you evaluate the current design approaches that are provided by the literature?**

**MW:** The literature on judgment and decision making and that on behavioral economics offers frameworks on how to design nudges. Information systems can certainly lean on these frameworks to create digital versions of nudges (e.g., the behavior change technique taxonomy, NUDGE, MINDSPACE, and the tools of a choice architecture), so I think we need fewer guidelines on how to create nudges (although such guidelines are certainly important) and more guidelines on how to test these nudges' effectiveness.

I am talking about the rigorous application of evidence-based practices like randomized controlled trial (RCTs) and their transparent and complete reporting. Only if results on the effectiveness of certain digital nudges are clearly reported can other authors take them up again, replicate them, and develop them further to create a knowledge body regarding nudging.

**AV:** I do not think there will ever be sufficient understanding. We live in a changing and evolving world, where new technologies and ideas appear all the time. There will always be insufficient understanding and always be ways for improvement. Not to mention that entirely new digital technologies may yet appear, which we will again have to explore, understand and learn to use. The design process itself, in its most general form, explore-understand-invent-build-test-repeat, will probably remain the same.

**MW:** I totally agree. The design process will remain quite similar; technologies and evidence will change, so we will never have sufficient understanding. However, in order to advance our understanding, we should, in the sense of evidence-based research, continue to test digital nudges in a rigorous scientific process to improve our understanding.

**CM:** **When we talk about building and testing such digital nudges, inspired by user-centered design and if we think about the overriding goal of nudging to improve the individual's decision behavior as judged by him- or herself, should such interventions be "user-centered digital nudges", including the involvement of users in the design process? Or is there another way in this regard?**

**AV:** Coming from the participatory modeling trade, I am certainly a big proponent of user participation in design. This participation is essential, especially when we are working on issues that are to impact human behavior and which are so intimately dependent on human perceptions, biases, culture and values. Yes, there is the famous quote from Steve Jobs that "A lot of times, people don't know what they want until you show it to them", and probably some geniuses can come up with superb products with no requirement engineering, creating jewels themselves. But even those can be always improved and personalized based on user requirements, especially when we are dealing with "unconscious thinking", which is almost impossible to formalize and estimate upfront.

Personally, with all my admiration of Apple, especially the way it was a decade ago, I find some features of their equipment very annoying and know they could be improved, if they could only listen to their customers.

**MW:** Yes, it certainly makes sense to involve users. In some situations, for example when user involvement is not feasible (maybe due to cost or time constrains), designers may proceed without involving users—but they should create nudges based on existing evidence and definitely take the ethical aspect of nudging seriously (such as nudges should benefit the user). We developed a relatively simple digital nudge-design cycle that contains four steps: 1) define the goal, 2) understand the user, 3) design the nudge, 4) test the nudge, and start again. If designers want to involve users, doing so in the second step would be particularly useful to understand users' decision-making processes.

**AV:** I would certainly imagine that the user is to be also involved in step four. It would also be helpful to have the user involved in step one. Especially, if we want to make the nudge useful for the user.

I do have a problem though with this proposed nudge design cycle. It should be described differently, otherwise it contains a vicious circle. We describe the design of a nudge, but then in step three we design the nudge. If we apply the same cycle, then step three should start all over again with step one, then step two, and so on. Perhaps step three should be called "implement the nudge".

**MW:** Fair point. I've just described the phases of the cycle only rudimentarily. In our paper, we describe each phases in more detail (e.g., step three also contains the implementation). We have also suggested shortcuts between the phases.

### 3.3 Challenges of the Digital Nudging Concept

**CM:** **Both of you think that users should be involved in the design process or that their preferences should be captured if possible. Referring to the understanding of nudging as preference-shaped interventions according to Hausman and Welch (2010): How can it be ensured, that designers of digital nudges really consider or at least try to consider the individuals' preference? Which ways exist to capture the preferences, which would be theoretically necessary before designing each and every nudge? Are there domains or intervention situations where it is easier or more difficult to consider the individual's preferences?**

**MW:** It is probably not possible to ensure that digital nudges consider individuals' preferences in every case. However, compared to conventional nudges, digital nudges can take advantage of all of the benefits of Information Systems, so they are slightly more preference-shaping. Data and algorithms may be used to derive preferences and then to tailor nudges dynamically to individual situations. The more data about a user is available, the more individual preferences can be taken into account, although doing so involves a privacy tradeoff.

**AV:** How do we know individual preferences when often individuals themselves do not know them? Our preferences can form only after been asked. We may have no preferences a priori, only when asked, when we start thinking about it, we can formulate our preferences. It does not mean they do not exist in the subconscious, fast thinking domain. It is just that we do not formulate them upfront. I do not know how these preferences can be extracted upfront from the subconscious to consider them when designing nudges. Moreover, I do not think it is important to know them when designing. If I want you to drink Coca-Cola, does it matter whether you currently prefer beer or tea?

**MW:** Agree, most people don't even know their preferences…

**CM:** **The potential problem of lacking preference awareness leads us to a similar challenge. The concept of nudging is based on the assumption that humans have limited cognitive capacities. Reflected in dual-process theory, human cognition processes can be categorized as 1) automated and unconscious thinking or 2) reflective and slow thinking, according to Kahneman (2011). It is in this distinguished view of human cognition that researchers clarify the transparency of a nudge as one pivotal aspect of nudging in distinction to manipulation. But how can a designer of digital nudges anticipate when a user will be in an unconscious or reflective state of thinking to avoid manipulation? How can nudgers be supported to deal with this issue?**

**MW:** That's a difficult question. As Kahneman (2011) noted, the categorization of human thought into two distinct systems is highly simplified. For example, in 2013, Evans and Stanovich distinguished between Type 1 (intuitive) and Type 2 (reflective) processes, where thought processes can be placed on a continuum between the two extremes. Some nudges address more intuitive thinking, while others address more reflective thinking. I like Sunstein's (2017) paper—"Do People Like Nudges?"—where he revealed that people support nudges in general, but they prefer reflective nudges to intuitive nudges (although they don't completely reject intuitive nudges).

> The question of manipulation is certainly again an ethical one, or a question of when manipulation begins. Merriam-Webster defines manipulation as the act "to change by artful or unfair means so as to serve one's purpose". Given this definition, someone would have to assess whether a nudge is unfair on a case-by-case basis.

**AV:** I do not think that nudging and, hence, digital nudging is very distinct from manipulation. Why would we be even developing the nudges if it was not something in human behavior that we were trying to change? We seem to be afraid of admitting it, and always assume that manipulation is bad. Yet it is all around us. Turn on your TV, open your newspaper, do a Google search—there you are: commercials are all around us, manipulating us to buy what we don't need, go to places we don't want to, and like people whom we despise. Somehow, they do not care about users avoiding manipulation. They just do their job and succeed.

And what is "unfair" in this context? Technically speaking, any nudge is unfair because, on the one hand, we have the computer savvy and well-educated nudge designer and, on the other hand, the general population that is quite unaware that it is being nudged.

**MW:** We're influenced every day, agree. Whether it is manipulation ("serve one's purpose") or another kind of influence is often a matter of definition.

**AV:** True, but in our case, we are talking about an influence that has a purpose (change behavior). Why bother otherwise? I understand that the 'm' word has a bad connotation and we do not want to admit that we are also doing it. But maybe it's better to admit it ourselves rather than be exposed by others?

**MW:** But manipulation always sounds like it's against the will of a person; nudging is all about being for the best of a person ("as judged by themselves"; Thaler & Sunstein, 2008).

**AV:** I understand that our intent is always good. But how do we know what is "for the best of a person"? If it is "judged by themselves", then how do we know what the nudge should be for? How do we design a nudge, which purpose we don't know? I'd be rather talking about nudging being for the best of the society. And then perhaps manipulation for that purpose is also not such a bad thing. Of course, this does not solve the problem of deciding what is actually for the best of the society and what is not. But we do have a certain history of democracy and can borrow some methods there.

**CM:** **Picking up on Markus' interesting comment regarding the assessment of nudges on a case-to-case basis, nudging also emphasizes on the importance to maintain the individual's freedom of choice at all time and hence all cases. How do you two generally understand the term "freedom of choice"? Are there choice architectures or intervention situations, in which the designer of nudges may not be able to comply with this goal of digital nudging?**

**AV:** Freedom of choice is an idealistic category, which is largely an illusion. In reality our choice is never really free. We are always impacted by the objective and subjective reality in which we exist. We base our choices not on what we like, but also on what we can, what we can afford, what we have time for. Jumping down from a roof of a 20-story building is certainly an example of our free choice, but we hardly have many opportunities to exercise that freedom. Probably only once in a lifetime.

**MW:** Yes, there is no neutral choice architecture. The presentation of choices always will influence the decision-maker's choice. Of course, people should be free to make their own choices, yet designers almost always consciously or unconsciously influence the decision environment in making their own choices. Let me give an example. Let's assume that we have a choice set with three equivalent options. If we preselect the first option, many will choose it (default bias), while if we leave a free choice, many will choose the middle option (middle-option bias), and if we set a limit on the third option, many will choose this option (scarcity bias). In all situations, people have a free choice, but they decide differently in each situation. We have tested these effects in a crowdfunding context. When considering freedom of choice, designers should be aware of these effects, but often they cannot anticipate all possibilities.

**AV:** Technically speaking this is correct. But if you know how the choice experiment works, and, if you deliberately present the option that you want to be chosen as the default one, is this really freedom? If we put the information that may swing the decision in small print, knowing that most likely it will not be read, how is this not manipulation? How much of freedom is left? Isn't this only proof that free choice is indeed largely an illusion?

**MW:** By freedom of choice I mean that options are not deliberately prohibited, but that all options remain available.

**AV:** Yes, but aren't some of the options deliberately made "more available"?

**MW:** Of course, that's one idea of nudging. Some option may be more available but users can also opt out and choose another option.

**CM: Besides the capability to avoid manipulation, respect the freedom of choice and consider the individual's preferences, designers of digital nudges also need to act correspondingly. Is there a way to "nudge the nudger"?**

**MW:** Certainly, there should be control authorities that control nudgers, but perhaps ethical guidelines can also help to "nudge the nudger".

**AV:** Sounds good, but if seriously, I do not think the nudger should be nudged. Creating nudges is a conscious process. Acting upon nudges is not. Controls and ethical requirements are certainly required, but why should they be performed as nudges?

**CM: Are there further challenges or limitations of the concept of Digital Nudging such as regarding goals, assumptions, or other?**

**MW:** One limitation is that (digital) nudging is often based on a simple human concept, that a human being who constantly makes mistakes because of cognitive limitations can be exploited by means of intuitive, "non-educative" (Type 1) nudges that steer behavior. However, voices are being raised that contend it is better to rely on educative nudges (Type 2) or "boosts" that foster people's competence in making choices–see for example Hertwig and Grüne-Yanoff (2017)–than on mechanisms that influence behavior (Type 1). A challenge for systems designers might be to implement digital boosts in a digital environment.

**AV:** I'm very much in favor of educative nudges. Exploiting the non-educative, intuitive, instinctive nudges is risky because of their manipulative character, where it may be harder to maintain ethical standards and overall fairness of the process. Moreover, let's face it, Kahneman's book is a bestseller, more and more people are getting well aware of the choice architecture and they will be less likely to follow the classic nudges that are based on default options and similar things.

### 3.4 Outlook to Further Digital Nudging Research

**CM: Markus, from your point of view, is there an important part that is still missing in the existing literature on digital nudging?**

**MW:** We should further study how technology influences the effectiveness of nudges, particularly the question concerning whether traditional nudges also work in digital environments, which is not always the case. For example, it has been shown that defaults do not work as well in online environments as they do in offline environments, perhaps because people in online environments are confronted much more often with default options and so have become more cautious. Whether and how conventional nudges can be transferred to digital environments are interesting questions.

**CM: And what are the most urgent aspects that you think design science researchers and behavioral science researchers should work on in upcoming years?**

**MW:** That's a good question. What is required is complementary research. Design science requires the design artifact (in our case, the nudge) to be evaluated, not just built, and behavioral science requires behavior to be analyzed if a stimulus material/artifact (again, the nudge) has been built beforehand. There is much to work on in the next few years, some of which I have already mentioned, but important topics include how technology affects nudging (i.e., effectiveness of digital nudges), how to deal with the ethical aspects of nudging in online environments (e.g., data protection and privacy), and how to ensure qualitative, transparent reporting of the results so a cumulative body of knowledge can emerge.

**AV:** Unfortunately, as any new technology, nudging may have its adverse effects. The digitalization of our social interactions, our new life in the virtual world of social media, where direct human interaction is replaced by interaction over mobile devices, where artificial intelligence becomes part of our everyday life and also starts nudging us in ways that may turn out to be totally out of our

control – all these uncharted areas definitely beg for attention, research, and, perhaps, moderation or direct regulation.

**CM:** **Alexey, with your center at UTS, how can digital nudging help to elicit sustainable and "wise adaptive living"? How could studies in this field look like?**

AV: Creating consumerism was easy. In a way we just had to unleash the "wild", "primitive" instincts that we already had in our genes. It is no surprise that some of Freud's principles were used by Bernays, who is seen as one of creators of consumerism, and who also happened to be Freud's nephew. Szejnwald and Vergragt (2016) provide a good review of the history of consumerism. From consumerism to wellbeing: toward a cultural transition. It will be more difficult to reverse the trend, to figure out how to practice self-restrain, altruism, sharing, and caring.

The only hope is for new knowledge, new technologies, new priorities. Social media and digital nudging are the kind of technologies that bear promise of influencing our behavior in ways that may be draw us to wiser living. We could certainly learn much from how conventional advertising works and see how its methods may be used to distribute the alternative messages. After all we are also talking about advertisement. But this time we want to advertise behavior and choices that can reduce our destruction of the Earth's life-support systems, to improve the social cohesion and build sharing and helping communities.

**CM:** **Markus, as a last question, do you personally have a certain focus for the future application on digital nudges, methodologically or domain specific?**

MW: In terms of method, I am working on how researchers can report experimental studies (e.g., nudging studies) in such a way that peers can easily replicate them—transparent, clear, and complete. In terms of content, I am currently working on how digital platforms can use nudges in the context of sustainability (e.g., on mobility platforms) or charitable giving (e.g., donation-based crowdfunding) to promote sustainable behavior or increase the willingness to donate. Most people probably agree ("as judged by themselves"; Thaler and Sunstein, 2008) that sustainable behavior and philanthropic behavior are socially relevant and are behaviors worth promoting or "nudging".

**CM:** **Thank you both very much for this discussion and interesting insights.**

## 4 Discussion

Digital nudging has become an important topic in IS research. Indeed, researchers at various conferences and in top-tier journals (including the basket of eight) have discussed it. Given that scholars introduced digital nudging research to the IS community only a few years ago (Weinmann et al., 2016), we see it as notable progress and a sign of the IS community's interest that the topic has garnered so much attention. While initial publications focused on definitions and models to design digital nudges (e.g., Mirsch et al, 2018), other work has investigated implementing and testing such nudges (e.g., Schneider et al., 2017; Kretzer & Maedche, 2018; Kroll & Stieglitz, 2019), which not only reflects the increasing advancements that digital nudging research has made but also the permanent interdependence between traditional design science research, which provides utility, and behavioral science research, which provides truth through theoretical advancements (Hevner, March, Park, & Ram, 2004).

At the same time, as our discussion with Markus Weinmann with Alexey Voinov also shows, the literature has insufficiently discussed the different options to evaluate the originality of digital nudging research and its conceptual distinction from research on, for example, persuasive technology. For instance, one can understand digital nudging as a soft and indirect way to change the users' minds through altering digital contexts and, hence, their digital choice environments, while persuasion focuses on more directly influencing users' minds (Cohen, 2013). In this case, one could see digital nudging as complementary to persuasion. Yet, as Alexey Voinov pointed out, one cannot argue that changing minds and contexts are separable, which leads to understanding digital nudging as a subcategory of persuasion. Both views seem generally acceptable. Rather than continuing this exhaustive debate on persuasion versus nudging, we establish three general aspects as argumentative grounds for future research to come. For one, argued from a historical view on both concepts, persuasion and, consequently, persuasive technologies in behavioral economics precedes digital nudging research. Therefore, it is possible but not imperatively necessary that digital nudging research learned and may have adopted some persuasion mechanisms as a subcategory of the overall persuasive technology research field. Secondly, given this temporal interdependency, it seems plausible to argue that one can characterize all digital nudging tools as persuasive mechanisms given the broad definition of persuasion as a super category. However, not all mechanisms used to persuade users in virtual environments qualify as nudging mechanisms as they can include, for example, monetary incentives to influence decision making. Third, we conclude that, as Markus Weinmann slightly touched on in the discussion, the most predominant and distinct differentiation between digital nudging and persuasion may lie in nudging's ethical aspects. Reviewing the literature, we found that IS research has only started to investigate and discuss this important aspect (Lembcke, Engelbrecht, Brendel, & Kolbe, 2019a). We believe that contributing more ethic-related research on digital nudging, especially research that uses and evaluates ethical guidelines for designing digital nudges, will further help digital nudging research to strongly demarcate aspects in the concept.

In discussing the existing literature above, we also identified a gap between what constitutes digital nudging and how one can actually put consequent requirements into practice, such as the requirement to increase users' welfare as they themselves judge it (Thaler & Sunstein, 2008). While the different design models of digital nudging emphasize the relevance of that goal (see e.g. Meske & Potthoff, 2017; Mirsch et al., 2018), they only vaguely discuss how to achieve it. At the same time, looking at existing empirical studies, it is hard to find any attempts to understand the users' preferences before altering their choice architecture or to even involve users in the design process. The same problem arises, for instance, when it comes to freedom of choice, which they should have according to the original nudging concept (Thaler & Sunstein, 2008). Partially questioning the general practicability of digital nudging, the discussion with Markus Weinmann and Alexey Voinov highlighted reasons for such issues, which lay in *inherent, methodical obstacles.* It is difficult to, for example, always know users' preferences, which they may not even know themselves in certain situations, or to consider all theoretically possible choices. Yet, nudgers also seem to lack awareness that altering digital choice environments only represents digital nudging if they consider and put its constituting characteristics into practice. Hence, we need a design science-oriented groundwork on how to overcome these issues. We postulate that the digital nudging research field—similar to what we have seen in the established design-oriented research (Hevner et al., 2004)—needs to carefully consider all aspects of the design process to assure rigid and generalizable digital nudging design processes that scholars can replicate in implementations in the future. Further, we found that much recent IS research on nudging focuses on developing and consequently empirically assessing digital nudges. However, future digital nudging research needs to complete known design-oriented research loops in combination with knowledge of what constitutes digital nudging in assessing what theory can learn from practice and vice versa (Hevner et al., 2004).

In addition, influencing users' behavior combined with a lack of knowledge about users' preferences may risk *manipulating* users *(*Hansen & Jespersen, 2013*).* The risk of manipulation also depends on a nudge's transparency and if it focuses on reflective or unreflective system of thinking. Should manipulation be undesirable, which our interview shows one can see differently, we again need *applicable ethical guidelines* for practitioners and researchers alike. Such guidelines could help regulate the overall situation in which computer-savvy and well-educated nudgers encounter a population that may not even aware they are being nudged. In this regard, nudgers could focus on educative over non-intuitive or instinctive nudges, which would help to maintain ethical standards and fairness. However, the differences between individual and societal preferences, which manifest in local or regional culture and traditions such as individualism versus collectivism (Hofstede, 2001), will significantly influence the degree to which digital nudges succeed, how scholars empirically evaluate them, and, consequently, the appropriateness of the digital nudging concept in general. Therefore, we suggest that we need to more deeply understand cultural differences when it comes to the effectiveness and appropriateness of digital nudging and the validity of the paternalistic paradigm in varying cultural settings.

Moreover, digital nudging research may investigate how technology itself influences the effectiveness of digital nudges since the online environment can lead to different perceptions of the choice architectures compared to those offline, which the literature has only partially discussed thus far (e.g., Hummel & Maedche, 2019). In consequence, validated nudges from the offline world may be ineffective in the online world. Also, the discussion highlights the need for additional approaches to test nudges and the relevance of qualitative and transparent reports of empirical results. Moreover, current literature indicates a gap of digital nudging research in(side) the corporate context, which calls for further investigations. However, as Alexey Voinov mentioned in the interview, we may never "perfectly" understand how to design and test digital nudges due to changes in our evolving world. Accordingly, artificial intelligence is a potential new actor in the nudging game, which the existing literature has not yet discussed.

## 5 Conclusion

In reviewing the literature, we found that digital nudging represents an important instrument to support the IS field's goal to increase human welfare (Fedorowicz, Bjørn-Andersen, Olbrich, Tarafdar, & Te'eni, 2019). Nevertheless, the discussion with Markus Weinmann and Alexey Voinov also showed that digital nudging research needs to further develop its self-conception (e.g., when does an intervention represent a digital nudge and, consequently, lead to an obligation to respect the paternalistic paradigm?) and instruments and guidelines (e.g., how can one put the obligations that the general digital nudging concept specifies into design practice?) to eventually help people advance towards wiser lifestyles and behavior.

Other fields, such as psychology, sociology, and computer science, have already tested various interventions of choice architectures in many different domains even if they have used different terms. While further investigating the effectiveness of nudges seems valuable to a certain extent, to increase the originality of IS research in this regard, the IS community could benefit from a stronger design science perspective to establish concrete design principles for different types of nudges and, hence, go beyond testing effectiveness in a specific case. We also need more conceptual and empirical research in the overall design process and corresponding obstacles, which needs to reflect on the definitory aspects that distinguish digital nudging from other approaches to influence user behavior. In this regard, concrete legitimizing conditions and the integration of techniques from user-centered or participatory design models could support IS nudgers to enhance results and extend the existing interdisciplinary body of knowledge.

## About the Authors

**Christian Meske** is Assistant Professor at the Department of Information Systems, Freie Universität Berlin, and Member of the Einstein Center Digital Future (Berlin), Germany. He was Coordinator of the Graduate School "User-Centered Social Media" at the Department of Computer Science and Applied Cognitive Science at University of Duisburg-Essen. His research on digital collaboration and transformation has been published in journals such as Business & Information Systems Engineering, Business Process Management Journal, Information Systems Frontiers, Information Systems Management, Journal of Enterprise Information Management, Journal of the Association for Information Science and Technology and various others. Amongst others, he has been recognized with the AIS Best Information Systems Publication of the Year Award.

**Ireti Amojo** is Research Assistant and PhD Candidate at the Department of Information Systems, Freie Universität Berlin, Germany. Ireti is also member of the Einstein Center Digital Future and holds a scholarship from the Foundation of German Businesses. She received her B.A. in Organizational Communications at the Washington State University and her M.Sc. in Applied Cognitive and Media Science at University of Duisburg-Essen. Her research on Digital Nudging and Enterprise Social Bots has been published in conferences like the Australasian Conference on Information Systems or Hawaii International Conference on System Sciences.